\documentclass[cits]{PoS}

\title{Progress on four flavor QCD with the HISQ action}

\ShortTitle{Progress on four flavor QCD with the HISQ action}

\author{The MILC Collaboration:}
\author{
\speaker{A.~Bazavov},$^a$ C.~Bernard,$^b$ C.~DeTar,$^c$ W.~Freeman,$^a$
Steven Gottlieb,$^d$ U.M.~Heller,$^e$
J.E.~Hetrick,$^f$ J.~Laiho,$^b$ L.~Levkova,$^c$ J.~Osborn,$^g$ R.~Sugar,$^h$ 
D.~Toussaint,$^a$ and R.S.~Van de Water$^i$\\
\llap{$^a$}
        Department of Physics, University of Arizona\\ Tucson, AZ 85721, USA\\
	E-mail: \email{bazavov@physics.arizona.edu}\\
\llap{$^b$}
        Department of Physics, Washington University\\
St. Louis, MO 63130, USA\\
\llap{$^c$}
        Physics Department, University of Utah\\
Salt Lake City, UT 84112, USA\\
\llap{$^d$} Department of Physics, Indiana University\\
	 Bloomington, IN 47405, USA\\
\llap{$^e$}
        American Physical Society\\
 One Research Road, Box 9000, Ridge, NY 11961, USA\\
\llap{$^f$}
        Physics Department, University of the Pacific\\
 Stockton, CA 95211, USA\\
\llap{$^g$}
	Argonne Leadership Computing Facility, Argonne National Laboratory\\
	Argonne, IL 60439, USA\\
\llap{$^h$}
        Department of Physics, University of California\\
 Santa Barbara, CA 93106, USA\\
\llap{$^i$}
        Department of Physics, Brookhaven National Laboratory\\
	Upton, NY 11973, USA\\
}

\abstract{We describe recent progress on generation of gauge 
configurations using the Highly Improved Staggered Quark (HISQ) action
that was designed by the HPQCD/UKQCD collaboration.
The HISQ action requires two levels of smearing with a reunitarization
of the links before the second smearing.  
We describe how we deal with the occurrence of occasional large
forces arising from the reunitarization step.
The MILC collaboration is currently
generating ensembles with approximate
lattice spacings of 0.15, 0.12, 0.09, and 0.06 fm,
with the strange and charm quark masses close to their physical values and
the mass of the light quarks $m_l$ set to $0.2 m_s$.
We present recent results for pion taste splittings, light
hadron masses, the static potential, the $\eta_c$ dispersion relation and 
the topological susceptibility.}

\FullConference{The XXVII International Symposium on Lattice Field Theory - LAT2009\\
		 July 26-31 2009\\
		 Peking University, Beijing, China}

\begin{document}

\section{Introduction}

The MILC collaboration has been using the asqtad action~\cite{ASQTAD} 
for more than a decade for its generation of 2+1
flavor dynamical quark configurations.
Over forty ensembles covering six lattice spacings 
from $a\approx 0.18$ fm to 0.045 fm and a wide range of
quark masses have enabled us to explore both the continuum and
chiral limits \cite{RMP}.  

Recently,
the Highly Improved Staggered Quark (HISQ) \cite{Follana:2006rc} action 
was developed by the HPQCD/UKQCD collaboration.
This action requires two levels of smearing with a reunitarization
of the links before the second smearing.  
Although it is more expensive to simulate than the asqtad action, its
reduced taste symmetry breaking makes the HISQ action an attractive
possibility for future calculations.  It also allows us to
include a dynamical charm quark with reasonable accuracy at current lattice
spacings.
The goal of this initial work is to
generate several ensembles with a fixed ratio of light to strange
quark masses and to study the approach to the continuum limit.

Large forces arise during molecular
dynamics updating when the determinant of a link at the first level
of smearing is close to vanishing \cite{Bazavov:2009jc}.  
This problem is dealt with by a slight
change in the guiding action, the effect of which can be eliminated
via the accept/reject step at the end of each trajectory.

The MILC collaboration is currently
generating ensembles with lattice spacings of about
0.15, 0.12, 0.09, and 0.06 fm,
with the strange and charm quark masses near their physical values and
the mass of the light quarks $m_{l}$ set to $0.2 m_s$.
In the future, we plan to generate additional ensembles with 
lighter up and down quark masses.
We present recent results on pion taste splittings, light
hadron masses, 
the $\eta_c$ dispersion relation and the topological susceptibility.

\section{Implementation of the HISQ action} 
    
The HISQ action was developed 
to go beyond the asqtad action in its improvement of taste-symmetry.
Links in the Dirac operator are replaced by
\begin{equation}
U' = {\cal F}_2 {\cal U} {\cal F}_1 U 
\end{equation}
where 
smearing level 1 denoted ${\cal F}_1$ is Fat 7; 
reunitarization is denoted by ${\cal U}$; and
smearing level 2 denoted ${\cal F}_2$ is asq smearing.



To calculate the fermion force, we must take the derivative of the action with
respect to the
gauge link.  With multilevel smearing, we must use the chain rule 
\cite{Kamleh:2004xk,Wong:2007uz}.  
For the fermion force, we have:
\begin{equation}
  \frac{\partial S_f}{\partial U}=
  \frac{\partial S_f}{\partial X}\,\frac{\partial X}{\partial W}\,
  \frac{\partial W}{\partial V}\,\frac{\partial V}{\partial U}\ ,
\end{equation}
where
$U$ are the fundamental gauge links, $V$ represent fat links after 
level 1 smearing, $W$ are the result of reunitarizing $V$, and $X$ are
the links after asq smearing.
Conveniently, code for three of the derivatives is the same as for the
asqtad action (although the arguments may be different):
\begin{equation}
  \frac{\partial S_f}{\partial X}\ ,\,\,\frac{\partial X}{\partial W}\ ,\,\,
  \frac{\partial V}{\partial U}\   .
\end{equation}


After the first level of smearing, the link $V$ must be projected into $U(3)$.
This required some new code.  We define:
$W=VQ^{-1/2}$, with $Q=V^\dagger V$.
The method to calculate the inverse square root, based on 
the Cayley-Hamilton theorem, had already been
applied by Morningstar \& Peardon \cite{Morningstar:2003gk}, and 
Hasenfratz, Hoffman \& Shaefer \cite{Hasenfratz:2007rf}.

\begin{equation}
Q^{-1/2}=f_0+f_1Q+f_2Q^2 \,. 
\end{equation}
We just need to find the eigenvalues of $Q$.

\begin{figure}
\begin{center}
\includegraphics[width=0.6\textwidth]{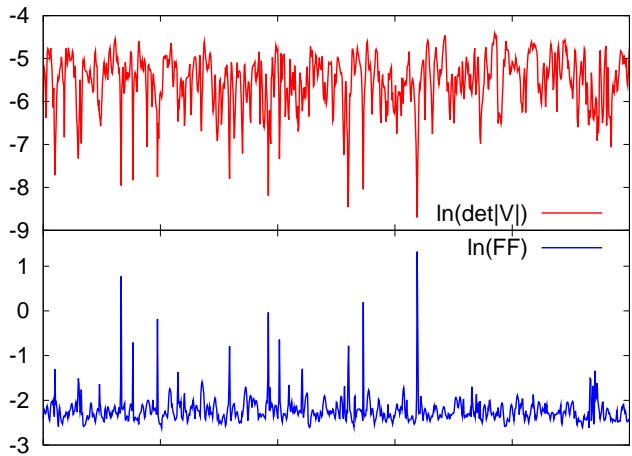}
\end{center}

\caption{Time history of the minimum value of
$\det|V|$ (upper plot) and the maximum magnitude of
the fermion force (lower plot) \cite{Bazavov:2009jc}.  
Note how the large spikes in the fermion
force are correlated with near vanishing of $\det|V|$.
The lattice spacing is $\approx 0.13$ fm.
\label{fig:force1}
}
\end{figure}
\begin{figure}
\begin{center}
\includegraphics[width=0.6\textwidth]{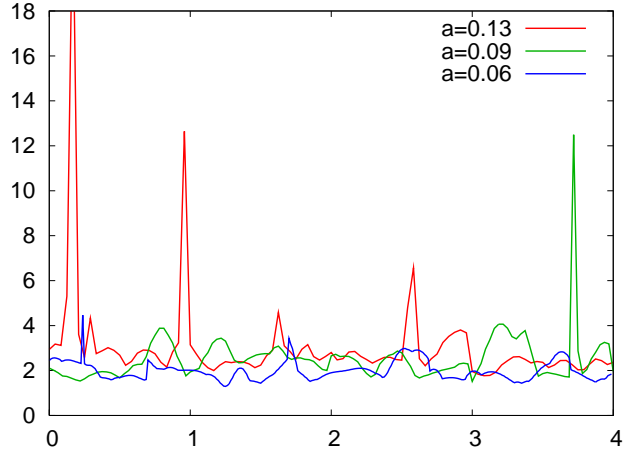}
\end{center}

\caption{Time history of maximum value of the fermion force rescaled 
by the inverse time step
for exploratory runs with three different lattice spacings.  See
Ref.~\cite{Bazavov:2009jc} for run parameters.
}
\label{fig:force2}
\end{figure}

A significant issue in implementing molecular dynamics based on this
algorithm is that there can be large changes in the action resulting
from large forces during the integration of the equations of 
motion \cite{Bazavov:2009jc}.
In Fig.~\ref{fig:force1}, we see that there is a correlation between 
the large forces and low values of $\det|V|$.
In Fig.~\ref{fig:force2}, we show the fermion force rescaled 
by $1/\Delta t$ for runs
with three different lattice spacings.  
(These two graphs are based on
exploratory runs details of which can be found in Ref.~\cite{Bazavov:2009jc}.)
It appears that as we approach
the continuum, there are fewer large forces.  As the fundamental links $U$
are varying more smoothly, we should not be surprised that $V$ is
less likely to have a small determinant.  


We implement an ``eigenvalue filter'' to tame the large forces.
If the smallest eigenvalue of
  a smeared link $V$ is smaller than a certain cutoff, we make the
replacement
  $ Q^{-1/2}\rightarrow(Q+\delta I)^{-1/2}$,
  where $\delta$ is typically set to $5\times 10^{-5}$.
Although this modifies the guiding Hamiltonian, the integration algorithm
is still reversible and area preserving.
As long as we use the RHMC algorithm \cite{Clark:2006fx}, 
the accept/reject step at the end of 
each MD trajectory ensures the correct equilibrium distribution.

\section{Testing the HISQ action}

For our initial test of the HISQ action, we are generating 
ensembles with $m_l = 0.2 m_s$ and lattice spacings varying from
$a\approx 0.15$ to 0.06 fm.
Table~1 shows the masses, lattice dimensions and
acceptance rate on our tuning runs.  In Table~2,
we consider the production runs at the two intermediate lattice spacings.
This range of lattice spacings is comparable to what we have 
with the asqtad action~\cite{RMP}
and we compare results for a number of quantities.
We find that the pion taste splittings are reduced by a factor of
2.5 to 3, and that results with the HISQ action are comparable to
those using the aqstad action at a lattice spacing of 
about 2/3 that used for the HISQ action.
\begin{table}
\begin{center}
\begin{tabular}{|l|l|l|l|l|l|l|} 
\hline
$\beta$ & $am_l$ & $am_s$ & $am_c$ & size & $a$ (fm) & acpt \\
\hline
5.80 & 0.0130 & 0.0650 & 0.838 & $16^3\times 48$ & 0.15 & 0.72 \\
6.00 & 0.0102 & 0.0509 & 0.635 & $20^3\times 64$ & 0.12 & 0.68 \\
6.30 & 0.0076 & 0.0380 & 0.445 & $28^3\times 96$ & 0.09 & 0.78 \\
6.65 & 0.0054 & 0.0270 & 0.313 & $48^3\times 144$ & 0.06& 0.70 \\
\hline
\end{tabular}
\end{center}
\caption{Parameters for tuning runs: coupling, dynamical quark masses
in lattice units, lattice dimensions, approximate lattice spacing, and
acceptance rate.}
\label{tab:runtablea}
\end{table}

\begin{table}
\begin{center}
\begin{tabular}{|l|l|l|l|l|l|l|}
\hline
$\beta$ & $am_l$ & $am_s$ & $am_c$ & size & $a$ (fm) & acpt \\
\hline
6.00 & 0.0102 & 0.0509 & 0.635 & $24^3\times 64$ & 0.12 & 0.74 \\
6.30 & 0.0074 & 0.0370 & 0.440 & $32^3\times 96$ & 0.09 & 0.77 \\
\hline
\end{tabular}
\end{center}
\caption{Same as for Table~1, except for production runs.}
\label{tab:runtableb}
\end{table}

\section{Initial results}
We present some initial results for the taste splittings of the pion,
the spectrum of particles with light quarks, the $\eta_c$ dispersion
relation and the topological susceptibility.  In most cases, the results are
compared with prior work on asqtad configurations and it is found that
the results with the HISQ action look closer to the continuum limit.  
(We do not have
results for the $\eta_c$ dispersion relation with asqtad quarks.)

\subsection{Pion taste splittings}

\begin{figure}[tbh]
\begin{center}
\includegraphics[width=0.55\textwidth]{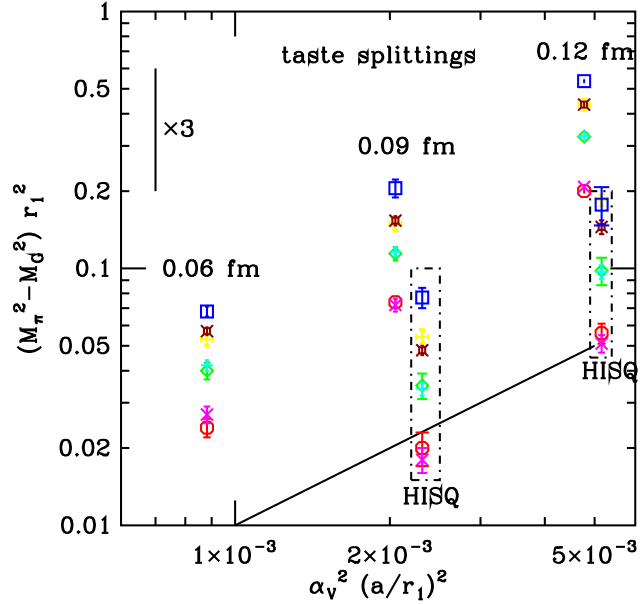}
\end{center}
                                                                                
\caption{Calculation of taste splitting using the
asqtad and HISQ actions.  We show
$(M_\pi^2-M_G^2)r_1^2$
{\em vs}. $\alpha_V^2 (a/r_1)^2$, where $M_G$ is the Goldstone taste state.
The diagonal line on this log-log plot shows the slope of the expected
dependence of the taste splitting.
}
\label{fig:tastesplit}
\end{figure}

It is well known that for staggered quarks, taste symmetry breaking is
an important lattice artifact.  A prime motivation for HISQ quarks is
to improve the taste symmetry.
In Fig.~\ref{fig:tastesplit}, we compare the pion taste splittings for the 
aqstad \cite{Bazavov:2009jc} and HISQ actions.  
We see that for $a\approx 0.12$ fm, the taste splittings for the HISQ action
are somewhat smaller than for asqtad with $a\approx 0.09$ fm.  Similarly,
results using the HISQ action with 
$a\approx 0.09$ fm are similar to results using the asqtad action
with $a\approx 0.06$ fm.
The vertical line to the left of this semi-log plot marked $\times3$ is a
ruler showing a factor of 3.  We see that using the HISQ action
reduces the taste splitting by a factor of 2.5 to 3 compared with asqtad.

\begin{figure}[hbt]
\begin{center}
\includegraphics[width=0.55\textwidth]{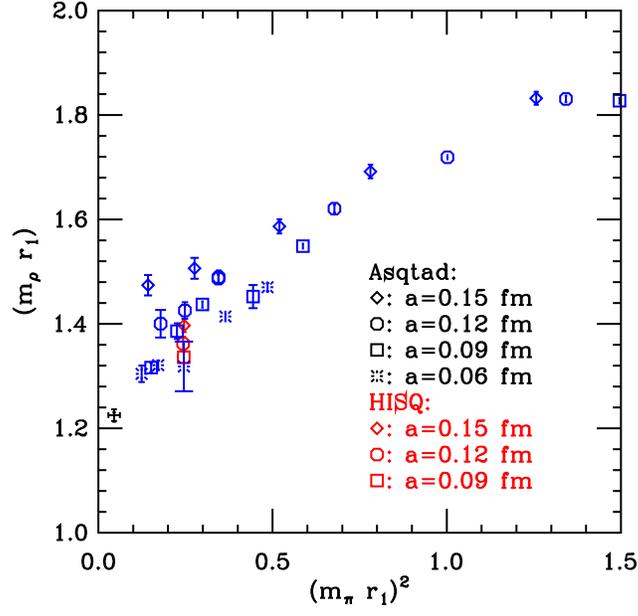}
\end{center}
\caption{Rho mass in units of $r_1$ {\em vs}. the pion mass
squared.
}
\label{fig:rho}
\end{figure}

\subsection{Spectrum of other light hadrons}

\begin{figure}
\begin{center}
\begin{tabular}{c c}
\includegraphics[width=0.45\textwidth]{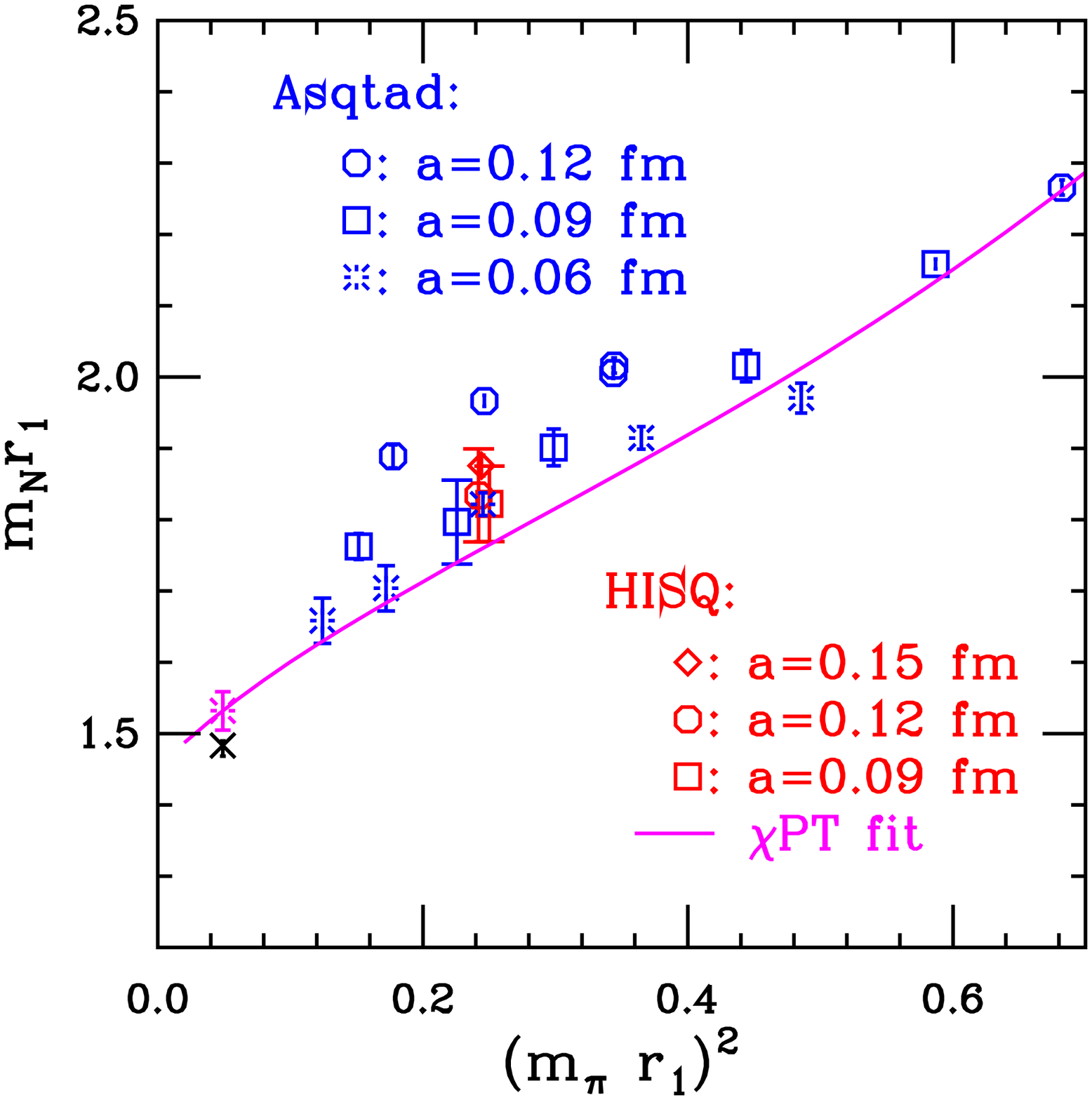}
&
\includegraphics[width=0.45\textwidth]{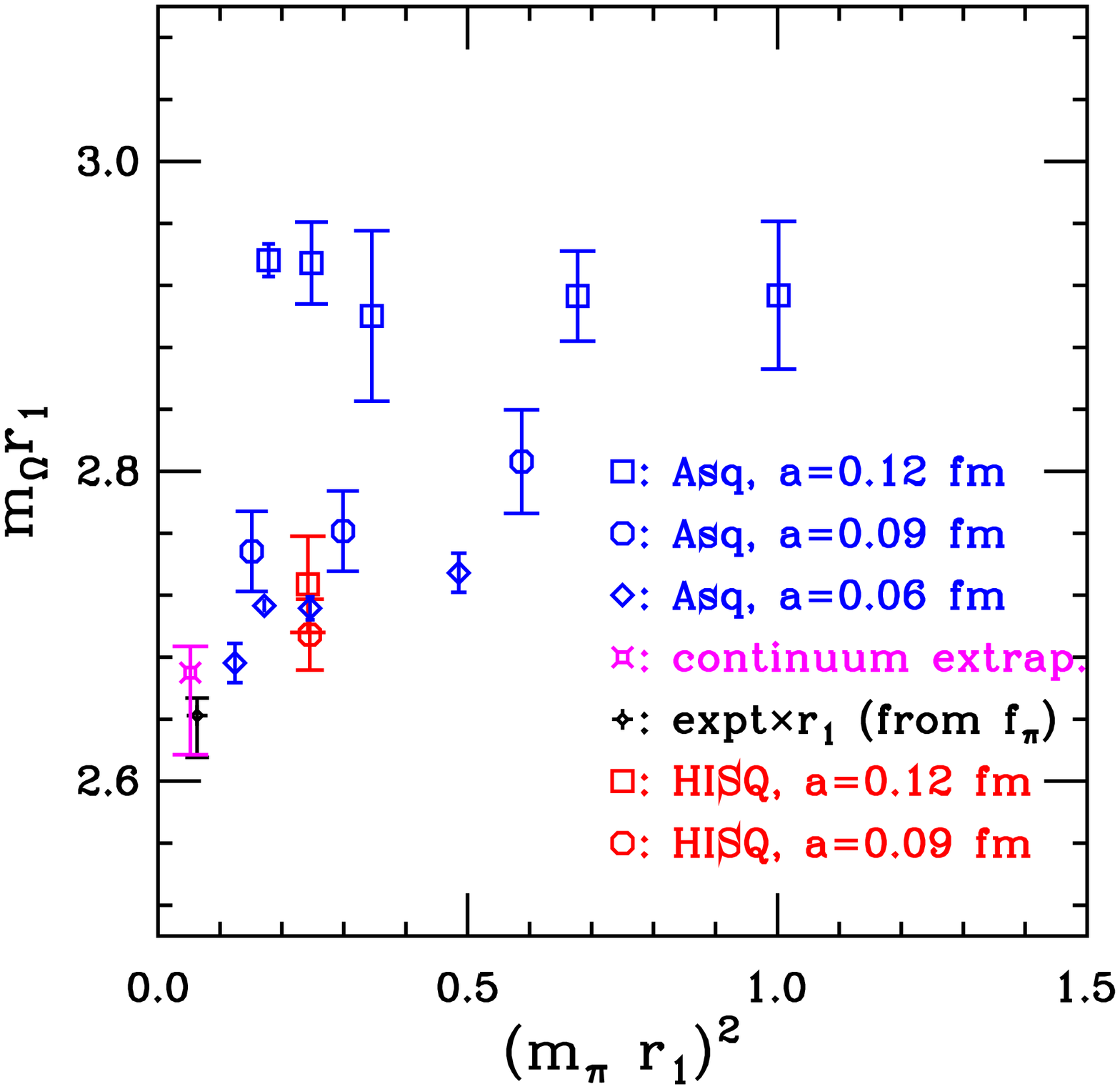}
\end{tabular}
\end{center}
\caption{Nucleon mass (left) and $\Omega$ mass (right) in units of $r_1$ 
{\em vs}.\ the pion mass squared.}
\label{fig:spectrum}
\end{figure}

We also have some preliminary results for the masses of $\rho$, nucleon
and $\Omega^-$.  In each case, we find that using the HISQ action
results in masses
with smaller lattice spacing dependence, and the results are comparable
to asqtad results at smaller lattice spacings.  (The poster presented
at the conference did not show results for $\rho$ due to lack of space.)
In Figs.~\ref{fig:rho} and \ref{fig:spectrum}, we show results for the $\rho$,
nucleon and $\Omega^-$.
The improvement for the $\rho$ seems quite substantial.
For the nucleon, we find that
the HISQ action result for $a\approx 0.15$ fm is quite a bit below the asqtad
result with $a\approx 0.12$ fm.  The HISQ result for $a\approx 0.09$ fm
appears comparable to that for the asqtad action
for $a\approx 0.06$ fm.  The magenta
curve represents a chiral perturbation theory fit to the asqtad results.
For the $\Omega^-$ we also see a substantial improvement of HISQ action results
over asqtad action results with $a\approx 0.12$ and 0.09 fm.

%

\subsection{HISQ action and the charm quark}
The HPQCD/UKQCD collaboration
has developed a procedure for tuning the coefficient of the Naik
term to reduce discretization errors up to ${\cal O}((am_c)^4)$.
At tree level, the correction to the Naik term expressed in terms of
the bare charm quark mass is \cite{Follana:2006rc}
\begin{equation}
\varepsilon = -\frac{27}{40} (am_c)^2 + \frac{327}{1120} (am_c)^4
       -\frac{15607}{268800} (am_c)^6 - \frac{73697}{3942400} (am_c)^8 \ .
\end{equation}
Note that this formula differs from Eq.~(24) in Ref.~\cite{Follana:2006rc}
because it is expressed in terms of the bare mass appearing in the action.
In Fig.~6, we show the dispersion relation for the $\eta_c$ meson.
For the three lattice spacings displayed, $-0.355 \le \epsilon \le -0.123$.
For $a\approx 0.06$ and $m_l=0.2 m_s$, $\epsilon \approx -0.063 $.
We see that for a lattice spacing as coarse as 0.15 fm, the deviation
from the continuum dispersion relation is as large as 7\%; however,
when the lattice spacing is reduced to 0.09 fm there is less than a 2\%
deviation from the continuum for the range of momentum shown here.

\begin{figure}[thb]
\begin{center}
\includegraphics[width=0.55\textwidth]{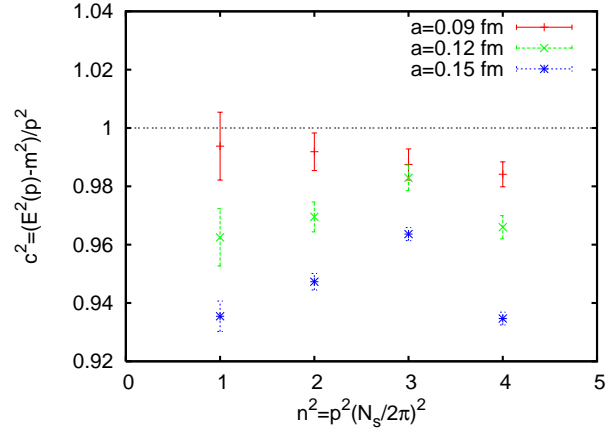}
\end{center}
\caption{Speed of light from the $\eta_c$ dispersion relation {\em vs}.
lattice momentum squared.
}
\label{fig:charm}
\end{figure}


\subsection{Topological Susceptibility}

In Fig.~7, we plot the topological susceptibility for both asqtad and
HISQ action simulations.  We find that the HISQ action point for 
$a\approx 0.12$ fm falls
close to the asqtad action curve for $a\approx 0.09$ fm.
Also note that the HISQ action point lies far to the left of the corresponding
asqtad action $a\approx 0.12$ point (the third green square from the left) 
since the HISQ action taste symmetry is so much better 
than for the asqtad action, and the horizontal axis involves
the taste singlet pion mass as shown by Billeter, DeTar 
and Osborn \cite{Billeter:2004wx}.
The black points and curve are from a continuum extrapolation of the fits to 
the results for the three asqtad lattice spacings.

\begin{figure}
\begin{center}
\includegraphics[width=0.55\textwidth]{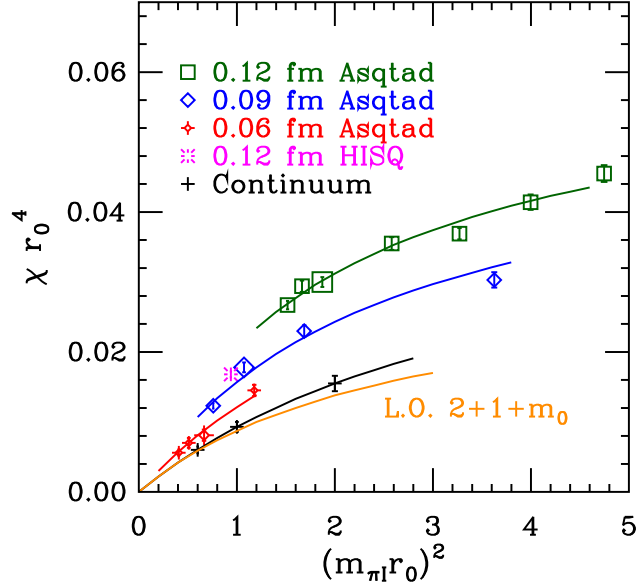}
\end{center}
\caption{ Topological susceptibility {\em vs}.\ squared mass of 
the taste singlet pion.  The symbol size has been increased for the
points coming from the asqtad action that correspond to $m_l=0.2 m_s$ so
that they can more easily be compared with the HISQ action point.  These
points are either the second or third point from the left, depending
on the lattice spacing.
}
\label{fig:topology}
\end{figure}
\section{Outlook}

Based on the encouraging results of our preliminary scaling study with
$m_l=0.2 m_s$, the MILC collaboration
is planning to generate additional ensembles with
$m_l=0.1 m_s$ and $0.04 m_s$ for lattice spacings between 0.06 and 0.15 fm.
These ensembles will allow us to study both the chiral and continuum limits.
We expect to be able to study a wide variety of physics with these 
ensembles, for example,
light pseudoscalar masses and decay constants,
light quark spectroscopy,
topology,
spectroscopy of particles with a heavy quark,
heavy-light meson decay constants and semileptonic decays, and
quark masses.

\section*{Acknowledgements}
This work was supported by the U.S.\ Department of Energy and the National
Science Foundation.  Computations were done on USQCD resources and at
NICS and TACC.

\end{document}